\documentclass[a4paper,11pt]{article}
\usepackage{pos}

\title{Searches for Neutrinos from Precursors and Afterglows of Gamma-Ray Bursts using the IceCube Neutrino Observatory}
 \ShortTitle{Search for neutrinos from GRBs}

\author{The IceCube Collaboration \\{\normalsize \normalfont(a complete list of authors can be found at the end of the proceedings)}}

\emailAdd{kunal.deoskar@fysik.su.se}
\emailAdd{paulcppn@gmail.com} \emailAdd{efriedman09@gmail.com}

\abstract{Gamma-ray bursts (GRBs) are among the most powerful events observed in our universe and have long been considered as possible sources of ultra-high-energy cosmic rays, which makes them promising neutrino source candidates. Previous IceCube searches for neutrino correlations with GRBs focused on the prompt (main emission) phase of the GRB and found no significant correlation between neutrino events and the observed GRBs. This motivates us to extend our search beyond the prompt phase. We perform analyses looking for evidence of neutrino emission up to 14 days before and after the prompt phase of GRBs. These analyses rely on a sample of candidate muon-neutrino events observed by IceCube from May 2011 to October 2018. The analyses are model-independent. Two of them scan different time-windows for possible neutrino emission, while a third analysis targets precursor emission based on GRB precursor observations by Fermi-GBM. We discuss the results and implications of these searches including limits on the contribution of GRBs to the diffuse neutrino flux.

\vspace{4mm}
{\bfseries Corresponding authors:}
Kunal Deoskar$^{1*}$, Paul Coppin$^{2}$,  Elizabeth Friedman$^{3}$\\

{$^{1}$ \itshape The Oskar Klein Centre, Department of Physics, Stockholm University}\\
{$^{2}$ \itshape Inter-University Institute For High Energies, Vrije Universiteit Brussel}\\
{$^{3}$ \itshape University of Maryland}\\[4mm]
$^*$ Presenter
}
\FullConference{37$^{\rm{th}}$ International Cosmic Ray Conference (ICRC 2021)\\
		July 12th -- 23rd, 2021\\
		Online -- Berlin, Germany}

\begin{document}
\maketitle
\section{Introduction}\label{sec:intro}


Gamma-ray bursts (GRBs) are bursts of intense gamma radiation, lasting from several ms up to several 100~s. GRBs have a non-thermal spectrum and have three phases of emission: precursor, prompt and afterglow. Based on the duration of the prompt phase ($T_{90}$\footnote{$T_{90}$ is the time interval that covers the central 90\% of the total gamma-ray photon count, starting at 5\% and ending at 95\%. Note that this parameter depends on the energy range of the observing detector.}), they are hypothesized to occur either due to the core collapse of super-massive stars (for $T_{90}$ > 2~s)  or due to the binary merger of compact objects (for $T_{90}$ < 2~s). Due to the extreme environments in which they occur, GRBs are potential sites of ultra-high-energy cosmic-ray production and therefore potential neutrino source candidates \cite{Neutrino}. IceCube analyses have been performed previously to search for neutrino emission from the prompt phases of GRBs and found no significant excesses \cite{IceCubeGRB2012,IceCubeGRB2015,IceCubeGRB2016,IceCubeGRB2017}. However, after the recent discovery of GRB activity following gravitational waves (GWs) from compact mergers by LIGO and Virgo~\cite{GW} as well as observations of gamma radiation well outside the prompt phase \cite{Precursors}, we are motivated to extend our search beyond the prompt phase to investigate a possible correlation between GRBs and the neutrinos detected by IceCube.

The IceCube Neutrino Observatory is a cubic-kilometer scale neutrino detection facility at the Geographic South Pole in Antarctica. It is a Cherenkov detector which makes use of 5160 Digital Optical Modules (DOMs) embedded in the glacial ice between the depths of 1450-2450 meters to detect astrophysical neutrinos.

Four analyses were designed to study the GRBs within the time frame of the IceCube neutrino-candidate events dataset and the results are presented in this proceeding. The ``Extended TW'' analysis, named for a whole range of different time windows, considers all GRBs, regardless of localization, within the data period.  The ``Precursor/Afterglow'' analysis focuses on well localized GRBs and performs a search separately for precursor and the prompt+afterglow phases without a fixed time window.  The ``GBM Precursor'' analysis focuses on precursor emission based on gamma-ray precursor observations by Fermi-GBM.  The ``Stacked Precursor'' analysis performs a stacked search on well-localized bursts and with a fixed time window.  The statistical significance of the population of results from the analyses is evaluated using the cumulative binomial probability and the post-trial final p-values for the four analyses are discussed respectively in this proceeding.

\section{Analysis method}\label{analysis_methods}
Each of the analyses makes use of an unbinned maximum likelihood method to quantify the potential correlation between GRB observations and IceCube events. In this section, we present details on the analysis method, GRB sample and IceCube events used in the analyses.

\subsection{GRB Catalog}
IceCube hosts a publicly available online GRB catalog\footnote{\href{https://icecube.wisc.edu/~grbweb_public/}{https://icecube.wisc.edu/~grbweb\_public/}}
that pools GRB observations from a wide range of GRB detectors and follow-up observatories. The GRB sample of the analyses share the same base list of GRBs from GRBweb, but each analysis imposes additional selection criteria on the GRB properties. These criteria, shown in Table \ref{tab:GRBs}, were motivated by their effects on the sensitivity and the compatibility of e.g. Fermi-GBM localizations with the analysis technique.

\begin{table}
    \caption{The number of GRBs used in each analysis, whether or not only GRBs with known $T_{90}$ durations are required, the maximum uncertainty on the localization, and the number of GRBs localized by Fermi-GBM.}
    \centering
    \begin{tabular}{|c|c|c|c|c|}
        \hline
        \textbf{Analysis} & \textbf{\# GRBs} & \textbf{Duration} &\textbf{Max Localization Uncertainty} & \textbf{\# GBM Localized} \\
        \hline
        Extended TW & 2091 & Required & --- & 1236 \\
        \hline
        Precursor/Afterglow & 733 & --- & 0.2$^\circ$ & --- \\
        \hline
        GBM Precursor & 133 & --- & --- & 100 \\
        \hline
        Stacked Precursor & 872 & --- & 1.5$^\circ$ & --- \\
        \hline
    \end{tabular}
    \label{tab:GRBs}
\end{table}

\subsection{IceCube event selection}
All analyses use the same sample of IceCube events that consists of well reconstructed muon tracks from May 2011 to October 2018. The vast majority of events that trigger the IceCube detector are not astrophysical neutrinos, but events related to cosmic-ray air showers. In the Southern hemisphere, atmospheric muons are observed at a rate of 2.7 kHz. Since only neutrinos can propagate through the Earth without being absorbed, this background vanishes in the Northern hemisphere, where atmospheric neutrinos dominate the background at the mHz level. A selection with different data quality cuts for the Northern and Southern hemisphere is therefore used, which reduces these backgrounds to 6.6 mHz integrated over the full sky. Assuming an $E^{-2}$ spectrum, the efficiency for selecting astrophysical neutrinos is estimated to be 95\% above 100~TeV in the Northern hemisphere and 70\% above 1~PeV in the Southern hemisphere. A detailed account of this event selection is given in \cite{GFUref}.

\subsection{Unbinned log likelihood / TS}

An unbinned log likelihood method is combined with frequentist statistics to assign a probability that a subset of neutrino candidate events is inconsistent with background.  The probability of each individual event being signal versus accidental coincidence is calculated based on its reconstructed energy, arrival time, and reconstructed direction relative to the GRB location.  Probability density functions (PDFs) are generated for the energy, space, and time of background (atmospheric muons and neutrinos) and signal (astrophysical neutrinos). 
%
%
The background energy PDF, $B(x_i)$, is constructed from data such that it represents the spectrum of all background events.  For the signal energy PDF, $S(x_i)$, the ``Precursor/Afterglow'' analysis allows the spectral index to be a fit parameter, while the other analyses assume an E$^{-2}$ spectrum. The background space PDF 
%
%
 is created from data and only varies as a function of declination as azimuthal symmetry can be assumed. The signal space PDF, shown in eq.~\eqref{sig_space_pdf}, uses a 2D Gaussian to determine the probability of the neutrino candidate's reconstructed position, $\vec{x}_\nu$, being consistent with the source position, $\vec{x}_{GRB}$,
\begin{equation}\label{sig_space_pdf}
    S(\vec{x}_\nu, \sigma | \vec{x}_{GRB}) = \frac{1}{2 \pi \sigma^2} exp \left( -\frac{|\vec{x}_\nu - \vec{x}_{GRB}|^2}{2 \sigma^2} \right)\ ,
\end{equation}
where $\sigma$ is the uncertainty on the reconstructed neutrino direction. Depending on the analysis method, the GRB position uncertainty is either negligible or taken into account as outlined below. The background time PDF is assumed to follow a uniform distribution across the analysis time window. The signal time PDF is a uniform box, which means that the PDF value is constant when the event falls within the time window and 0 when it falls outside the time window. The specific time windows used in each analysis are described in Section \ref{overview}.

The probability density functions 
are applied to every event $x_i$ in the dataset to obtain the following test statistic:
\begin{equation}
    TS = \ln \left[\frac{L(\hat{n}_s)}{L(n_s=0)} \right] = -\hat{n}_s + \sum^N_{i=1} \ln \left[\frac{\hat{n}_s S(x_i)}{\langle n_b \rangle B(x_i)} +1 \right]\ ,
    \label{eq:TS}
\end{equation}
where $\langle n_b \rangle$ is the average number of background events and $\hat{n}_s$ the number of fitted signal events that maximize the log likelihood ratio. High-energy events in temporal and close spatial coincidence will contribute most significantly to the test statistic.  The p-value of a given GRB is determined by comparing the unblinded test statistic to a test statistic distribution created from scrambled data.

The ``Extended TW'' and ``GBM Precursor'' analyses have additional steps to create the test statistic distribution due to the poor localization of GRBs solely detected by the Fermi-GBM satellite \cite{GBMUncertainty}.  All-sky test statistic maps of scrambled neutrino data are combined with a probability map provided by Fermi-GBM for a given GRB. The probability $P_\mathrm{GBM}$ acts as a penalty to the test statistic:
%
%
%
%
\begin{equation}
    TS_\mathrm{final} = TS_\mathrm{original} + 2\times [\ln(P_\mathrm{GBM}) - \ln(P_\mathrm{GBM,max})]\ ,
\end{equation}
where $TS_\mathrm{original}$ refers to the test statistic from Eq. (\ref{eq:TS}). The position of the GRB on the sky and the number of signal events are thus fitted, to find the combination which maximizes $TS_\mathrm{final}$.


\subsection{Cumulative Binomial Test}\label{binomial_test}
\begin{figure}[t]
    \centering
    \includegraphics[width=0.49\linewidth]{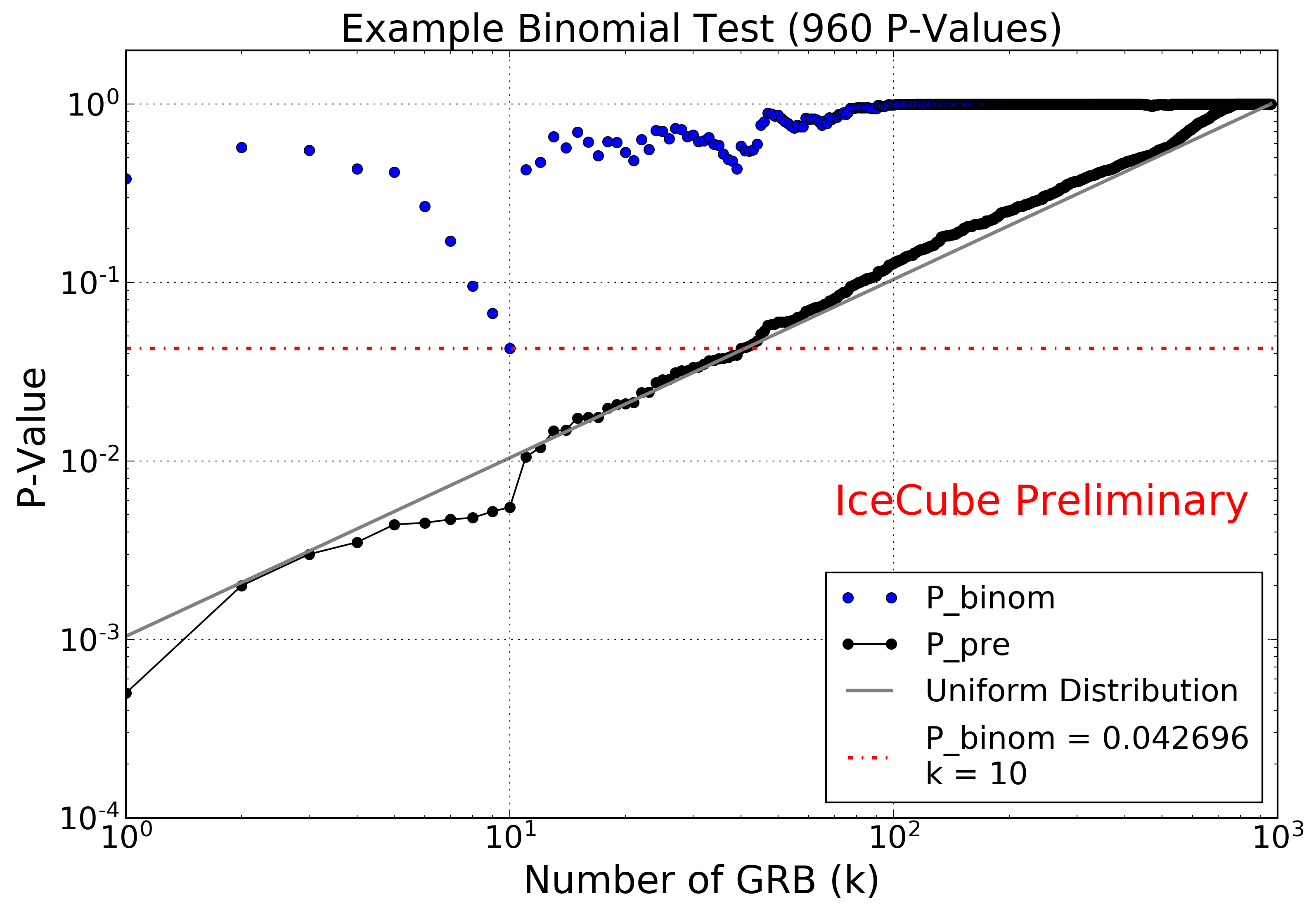}
    \includegraphics[width=0.49\linewidth]{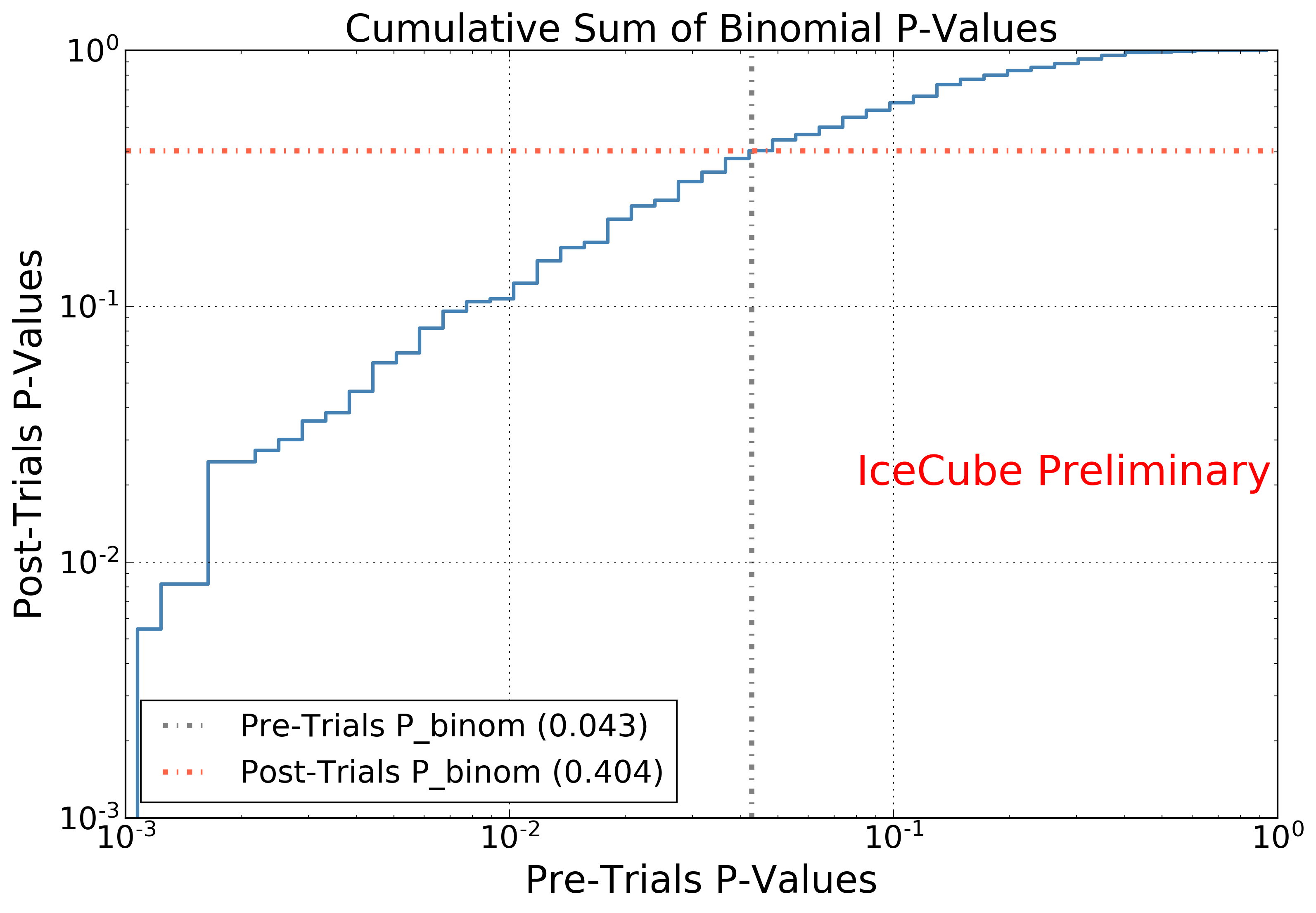}
    \caption{Left: An example of the binomial test based on simulation.  In this trial, k=10 was found to be most significant with a binomial p-value of 0.043 (red line).  Right: The post-trials p-value (red line) correction of the binomial p-value (blue line) found in the test on the left.  The cumulative sum in this plot is created from scrambled data, which includes possible correlations between GRBs.}
    \label{fig:binom_test_example}
\end{figure}

Analyzing a selection of $N$ GRBs will result in a $p$-value for each individual burst. A trials-correction method is thus needed to determine if a subset of the obtained $p$-values provides a statistically significant result. Arranging the unblinded $p$-values from smallest to largest, their values are denoted as $p_1,\ p_2, ..., p_N$. Under the background hypothesis $\left(n_s=0\right)$, these $N$ $p$-values are expected to follow a uniform distribution between 0 and 1. The probability to find $k$ or more $p$-values that are smaller than or equal to $p_k$ thus corresponds to the following binomial distribution:
\begin{equation}\label{binomial}
    P(k) \equiv P(n\geq k|N,p_k) = \sum_{m=k}^N \frac{N!}{(N-m)!m!} p_k^m (1-p_k)^{N-m}\ .
\end{equation}
Looping over all potential $k$, the smallest $P(k)$ is selected and used to define a test statistic $TS=-2\log\left(P(k)\right)$. A frequentist approach is then used to determine the significance of the observed $TS$ value based on the background $TS$-distribution. This binomial test is illustrated in Fig. \ref{fig:binom_test_example} and was used to determine the final $p$-value for the ``Extended TW'' and ``Precursor/Afterglow'' analyses.  It was verified that the unblinded result is not artificially boosted by a single IceCube event contributing to the significance of multiple GRBs.

A slightly different method was used in the ``GBM Precursor'' analysis. Instead of considering the $k$-th smallest $p$-value, the product of the $k$ smallest $p$-values was used, as this resulted in an improved sensitivity and discovery potential for that particular analysis.  

\subsection{Analysis approach}\label{overview}
\begin{figure}
    \centering
    \includegraphics[width=0.9\linewidth]{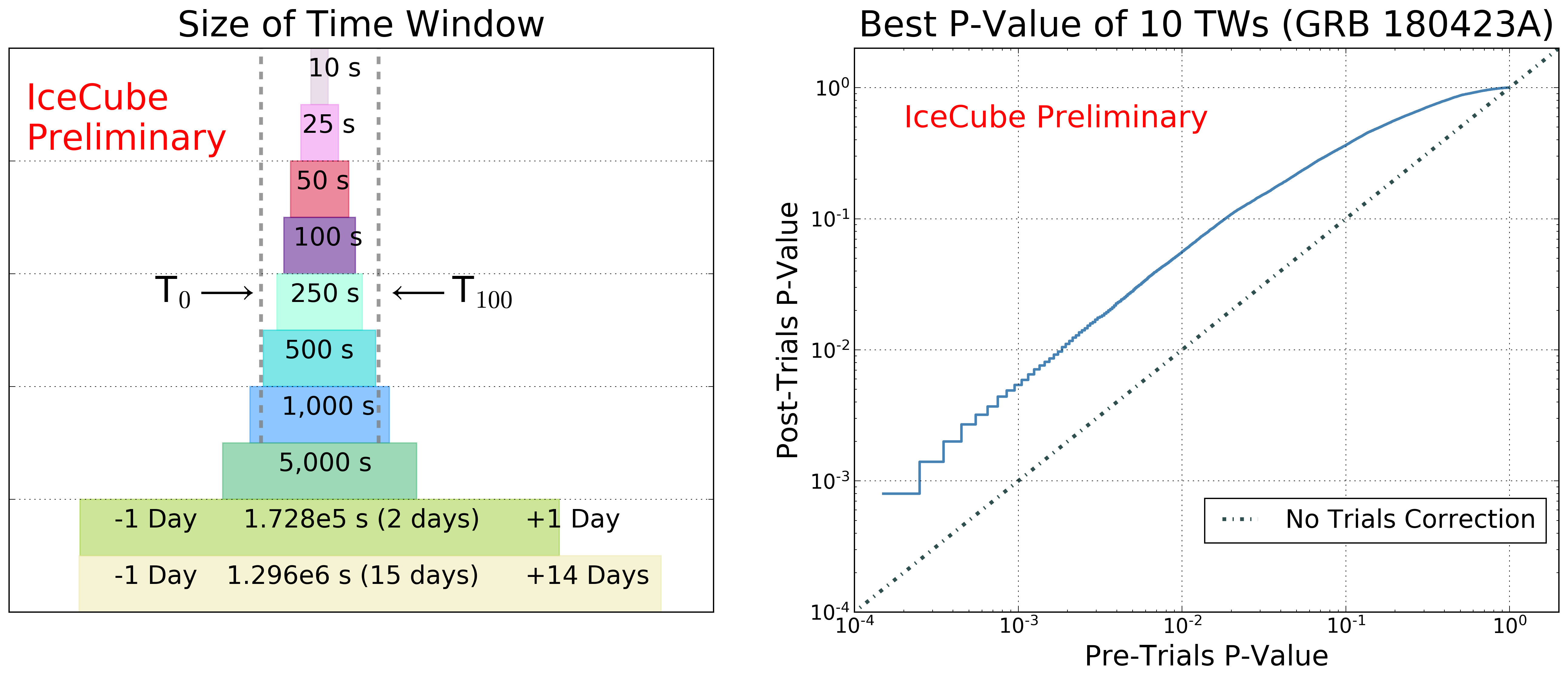}
    \caption{Left: The time windows used in the ``Extended TW'' analysis.  These 10 time windows are evaluated for all GRBs, regardless of their duration.  The first 9 time windows are centered on $T_{100}$, while the longest time window is asymmetric about the GRB duration.  Right: An example of an effective trials correction used to correct the most significant of 10 p-values to the final post-trials p-value for a given GRB.}
    \label{extended_tw_cartoon}
\end{figure}

\noindent\textbf{Extended TW. }
The ``Extended TW'' analysis calculates p-values for 2,091 GRBs in 10 pre-determined time windows and selects the most significant of those 10 p-values.  The pre-determined time windows range from 10 seconds to 2 days centered on the $T_{100}$\footnote{$T_{100}$ is defined as the time between the earliest and latest measured gamma-ray excess.} of the GRB, and the final time window is asymmetric with a 1 day precursor and 14 day afterglow (see Figure \ref{extended_tw_cartoon}).
%

The trials correction for searching 10 time windows is also shown in Figure \ref{extended_tw_cartoon}. To determine the effective trials correction for a given GRB, a set of pre-generated test statistic scans is used to calculate the background-only p-value in each time window.  The test statistic scans are carefully controlled to ensure that the scrambled data is the same in each time window.  The data in the 10 second time window is exactly contained in the 25 second time window, and so on.  The best of the 10 p-values for each scramble is added to a cumulative sum, which is used as a look-up table to correct the unblinded p-value for a given GRB.

The p-values that have been corrected for searching 10 time windows are then evaluated in a binomial test (Section \ref{binomial_test}) to correct for the population size.
\\

\noindent\textbf{Precursor/Afterglow.}
The ``Precursor/Afterglow'' analysis searched for neutrino excesses separately from the precursor phase and from the prompt+afterglow phase of GRBs. GRBs with positional angular uncertainty of less than 0.2\textdegree\ were considered for this analysis so that they can be approximated as point sources in the IceCube analysis. This resulted in a catalog of 733 GRBs and searches were performed separately for the precursor and afterglow phases of these GRBs. The best-fit values were obtained for the number of signal events ($n_s$), spectral index ($\gamma$), and time window of emission ($T_w$) for every GRB in each search. Regarding the temporal part of the Signal PDF, one end of the flat box shaped time window was assumed to be at the start of the prompt phase ($T_0$) and the other end was fitted using the data either up to 14 days prior to $T_0$ (for `precursor' searches) or up to 14 days after $T_0$ (for `prompt+afterglow' searches). Every GRB reported a p-value as well as the best-fit parameters for the respective search. This altogether resulted in two lists of 733 p-values, one for each search. The binomial test was performed on the two lists of results for the respective searches to investigate for statistically significant signals from a smaller population of sources from our GRB selection. The overall significance of each search is determined using the final post trial p-value which is obtained using the binomial test result for the respective search.

\begin{figure}
    \centering
    \includegraphics[width=0.60\linewidth]{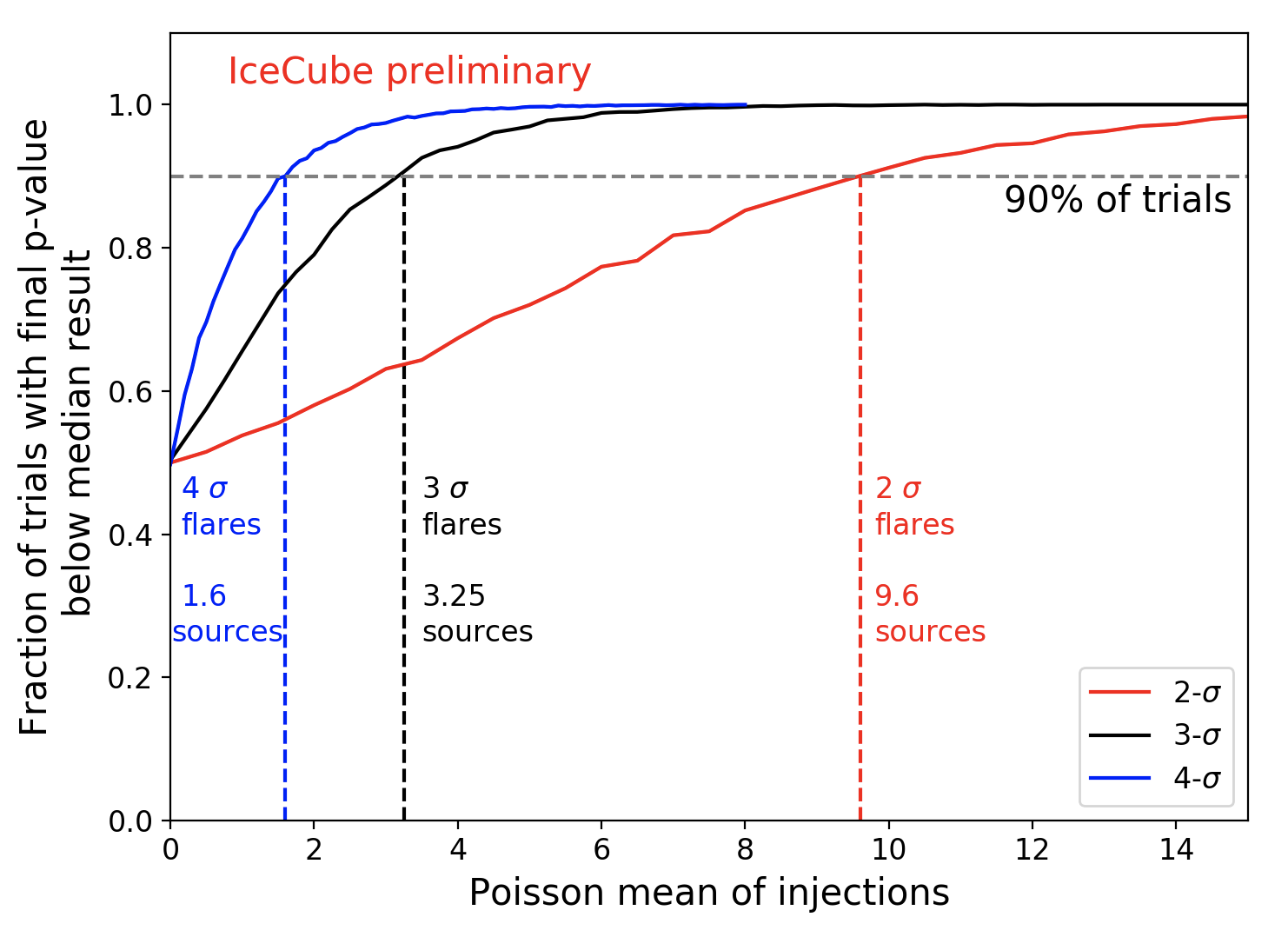}
    \caption{Post-trial sensitivity calculation for the ``Precursor/afterglow'' search. This is based on 10,000 instances of simulated datasets. Different combinations of individual p-values can lead to the same final p-value. As evident from our study, different numbers of individual n-$\sigma$ results will result in 90\% of our trials reporting a final p-value better than 0.5.}
    \label{sens_final_plot}
\end{figure}

We performed a study to test the signal detection sensitivity of our analysis (see Figure~\ref{sens_final_plot}). We define the sensitivity as the 90\% Confidence Level (CL) upper limit in the case that there is no signal. Hence it is the signal strength which, in 90\% of the cases, would give a lower final p-value than the median of the final background TS distribution. Figure~\ref{sens_final_plot} shows how different numbers of individual 2-$\sigma$,3-$\sigma$ and 4-$\sigma$ p-values can result in 90\% of our trials reporting a final post-trial p-value better than the median result. 
\\

\noindent\textbf{GBM Precursor \& Stacked Precursor.}
An analysis of the light curves of 2368 Fermi-GBM GRBs found evidence for precursor activity in $\sim$10\% of all GRBs \cite{Precursors}. Gamma-ray precursors indicate that central engine activity is already ongoing at an early stage and offer a precise time window to look for counterpart neutrino emission. A selection of GRBs with precursor emission was made \cite{Precursors}, resulting in 133 precursor bursts for which IceCube data is available. For each burst, the time window of precursor emission, extended by 2~s on either side, was searched for neutrino emission. The 133 p-values were then trial-corrected by considering the product of the $k$ most significant p-values.

The ``GBM Precursor'' analysis also revealed \cite{Precursors} that almost all (>95\%) precursors occur within 250~s of the prompt emission. A second analysis, looking for excess neutrino emission in the 250~s that precedes the start of the prompt phase, was therefore implemented. This second analysis used 872 well localised bursts (as indicated in Table \ref{tab:GRBs}) and a stacking procedure in which only a single $n_s$ value is fitted for the combined set of GRBs. This stacking approach results in an improved sensitivity compared to analysing each GRB individually, but restricts the analysis to well localized GRBs.

\section{Results}\label{sec:results}


\begin{table}
    \centering
    \caption{Post-trials p-value of the binomial test for the ``Extended TW'' study of 2,091 GRBs.  The binomial test was run on four subsets of GRBs split by hemisphere and duration.  The number of GRBs in each sub-population is indicated in parentheses.}
    \begin{tabular}{|c|c|c|c|}
        \hline
        Northern Long (960) & Northern Short (183) & Southern Long (814) & Southern Short (134) \\
        \hline
        0.038 & 0.799 & 0.898 & 0.849 \\
        \hline
    \end{tabular}
    \label{liz_binomial_results}
\end{table}

\begin{table}
    \centering
    \caption{Statistics for the top GRB for each sub-population in the ``Extended TW'' study of 2,091 GRBs.  The number of GRBs in the sub-population is indicated in parentheses.  The pre-trials p-value listed is corrected for searching ten time windows, but is not corrected for the population size.  The post trials p-value is corrected for the population size and for searching ten time windows.}
    \begin{tabular}{|c|c|c|c|c|}
        \hline
        GRB Name & Sub-Population & Most Significant Time Window & P$_{\mathrm{pre}}$ & P$_{\mathrm{post}}$ \\
        \hline
        140607A & Northern Long (960) & $\pm$1 Day & 6.0e-04 & 4.4e-01 \\
        \hline
        GRB140807500 & Northern Short (183) & 100 Seconds & 4.8e-03 & 5.9e-01 \\
        \hline
        150202A & Southern Long (814) & $\pm$1 Day & 5.0e-04 & 3.3e-01 \\
        \hline
        GRB140511095 & Southern Short (134) & $\pm$1 Day & 9.2e-03 & 7.1e-01 \\
        \hline
    \end{tabular}
    \label{liz_top_grbs}
\end{table}

\subsection{Extended TW}\label{liz_results} 

2,091 GRBs were evaluated for neutrino correlation in 10 time windows (Figure \ref{extended_tw_cartoon}), with the most significant p-value selected to represent the GRB.  Those 2,091 p-values are split into sub-populations by hemisphere and duration, and evaluated with a binomial test.  The results of the binomial test are summarized in Table \ref{liz_binomial_results} and are consistent with background.  The GRB with the most significant result from each sub-population is listed in Table \ref{liz_top_grbs}.

\subsection{Precursor/Afterglow}\label{Kunal} 

The 733 GRBs considered for this analysis were analysed individually for the two different searches. This resulted in one list of 733 individual p-values and best-fit parameters for the precursor search and another list of 733 individual p-values and best-fit parameters for the prompt+afterglow search. The binomial test was performed on each of these lists and the statistical significance of the best binomial p-values thus obtained are 0.495 for the precursor search and 0.486 for the prompt+afterglow search respectively. 
The results are comparable with a median background expectation (final p-value of 0.5).

\subsection{GBM Precursor \& Stacked Precursor}\label{Paul} 
Considering only events within a relative combined neutrino and GRB angular uncertainty of $5\sigma$, no neutrino events arrived in temporal coincidence with the precursors of the 133 GRBs. The second stacking analysis did find five IceCube events in temporal and spatial coincidence with the 872 GRBs. 
Given that a 250~s time window was examined per GRB and given that all five events have energy consistent with that of background events, this observation is fully consistent with the background expectation. Hence, this stacking analysis also resulted in a $p$-value of 1.


\section{Conclusion}

We searched in the direction of GRBs reported by various instruments to check for neutrino emissions not limited to the prompt phases of GRBs. The four analyses presented here all report observations consistent with background expectations. The binomial tests did not report any significant sub-populations with statistically significant results for the GRBs considered in each analysis. The results from these analyses will be used in future work to compute upper limits on neutrino fluxes from GRBs as well as to constrain the contribution to IceCube diffuse flux. These results can subsequently be used to constrain physical models such as that presented in~\cite{PrecursorModel}.

\bibliographystyle{ICRC}
\bibliography{references}

\providecommand{\href}[2]{#2}\begingroup\raggedright\begin{thebibliography}{10}

\bibitem{Neutrino}
K.~{Asano} and K.~{Murase} \href{http://dx.doi.org/10.1155/2015/568516}{{\em
  Adv. Astron.} {\bfseries 2015} (Jan., 2015) 568516}.

\bibitem{IceCubeGRB2012}
M.~G. {Aartsen} {\em et~al.} \href{http://dx.doi.org/10.1038/nature11068}{{\em
  Nature} {\bfseries 484} no.~7394, (Apr., 2012) 351--354}.

\bibitem{IceCubeGRB2015}
M.~G. {Aartsen} {\em et~al.}
  \href{http://dx.doi.org/10.1088/2041-8205/805/1/L5}{{\em ApJL} {\bfseries
  805} no.~1, (May, 2015) L5}.

\bibitem{IceCubeGRB2016}
M.~G. {Aartsen} {\em et~al.}
  \href{http://dx.doi.org/10.3847/0004-637X/824/2/115}{{\em ApJ} {\bfseries
  824} no.~2, (June, 2016) 115}.

\bibitem{IceCubeGRB2017}
M.~G. {Aartsen} {\em et~al.}
  \href{http://dx.doi.org/10.3847/1538-4357/aa7569}{{\em ApJ} {\bfseries 843}
  no.~2, (July, 2017) 112}.

\bibitem{GW}
B.~P. {Abbott} {\em et~al.}
  \href{http://dx.doi.org/10.3847/2041-8213/aa920c}{{\em ApJL} {\bfseries 848}
  no.~2, (Oct., 2017) L13}.

\bibitem{Precursors}
P.~{Coppin}, K.~D. {de Vries}, and N.~{van Eijndhoven}
  \href{http://dx.doi.org/10.1103/PhysRevD.102.103014}{{\em PRD} {\bfseries
  102} no.~10, (Nov., 2020) 103014}.

\bibitem{GFUref}
M.~G. {Aartsen} {\em et~al.}
  \href{http://dx.doi.org/10.1016/j.astropartphys.2017.05.002}{{\em
  Astroparticle Physics} {\bfseries 92} (June, 2017) 30--41}.

\bibitem{GBMUncertainty}
A.~{Goldstein} {\em et~al.}
  \href{http://dx.doi.org/10.3847/1538-4357/ab8bdb}{{\em ApJ} {\bfseries 895}
  no.~1, (May, 2020) }.

\bibitem{PrecursorModel}
S.~{Razzaque}, P.~{M{\'e}sz{\'a}ros}, and E.~{Waxman}
  \href{http://dx.doi.org/10.1103/PhysRevD.68.083001}{{\em PRD} {\bfseries 68}
  no.~8, (Oct., 2003) 083001}.

\end{thebibliography}\endgroup



\clearpage
\section*{Full Author List: IceCube Collaboration}




\scriptsize
\noindent
R. Abbasi$^{17}$,
M. Ackermann$^{59}$,
J. Adams$^{18}$,
J. A. Aguilar$^{12}$,
M. Ahlers$^{22}$,
M. Ahrens$^{50}$,
C. Alispach$^{28}$,
A. A. Alves Jr.$^{31}$,
N. M. Amin$^{42}$,
R. An$^{14}$
% ...
\\

\noindent
% ...
$^{12}$ Universit{\'e} Libre de Bruxelles, Science Faculty CP230, B-1050 Brussels, Belgium \\
% ...
$^{14}$ Department of Physics and Laboratory for Particle Physics and Cosmology, Harvard University, Cambridge, MA 02138, USA \\
% ...
$^{17}$ Department of Physics, Loyola University Chicago, Chicago, IL 60660, USA \\
$^{18}$ Dept. of Physics and Astronomy, University of Canterbury, Private Bag 4800, Christchurch, New Zealand \\
% ...
$^{22}$ Niels Bohr Institute, University of Copenhagen, DK-2100 Copenhagen, Denmark \\
% ...
$^{28}$ D{\'e}partement de physique nucl{\'e}aire et corpusculaire, Universit{\'e} de Gen{\`e}ve, CH-1211 Gen{\`e}ve, Switzerland \\
% ...
$^{31}$ Karlsruhe Institute of Technology, Institute for Astroparticle Physics, D-76021 Karlsruhe, Germany  \\
% ...
$^{42}$ Bartol Research Institute and Dept. of Physics and Astronomy, University of Delaware, Newark, DE 19716, USA \\
% ...
$^{50}$ Oskar Klein Centre and Dept. of Physics, Stockholm University, SE-10691 Stockholm, Sweden \\
% ...
$^{59}$ DESY, D-15738 Zeuthen, Germany \\
% ...

\subsection*{Acknowledgements}

\noindent
USA {\textendash} U.S. National Science Foundation-Office of Polar Programs,
U.S. National Science Foundation-Physics Division,
U.S. National Science Foundation-EPSCoR,
Wisconsin Alumni Research Foundation,
Center for High Throughput Computing (CHTC) at the University of Wisconsin{\textendash}Madison,
Open Science Grid (OSG),
Extreme Science and Engineering Discovery Environment (XSEDE),
Frontera computing project at the Texas Advanced Computing Center,
U.S. Department of Energy-National Energy Research Scientific Computing Center,
Particle astrophysics research computing center at the University of Maryland,
Institute for Cyber-Enabled Research at Michigan State University,
and Astroparticle physics computational facility at Marquette University;
Belgium {\textendash} Funds for Scientific Research (FRS-FNRS and FWO),
FWO Odysseus and Big Science programmes,
and Belgian Federal Science Policy Office (Belspo);
Germany {\textendash} Bundesministerium f{\"u}r Bildung und Forschung (BMBF),
Deutsche Forschungsgemeinschaft (DFG),
Helmholtz Alliance for Astroparticle Physics (HAP),
Initiative and Networking Fund of the Helmholtz Association,
Deutsches Elektronen Synchrotron (DESY),
and High Performance Computing cluster of the RWTH Aachen;
Sweden {\textendash} Swedish Research Council,
Swedish Polar Research Secretariat,
Swedish National Infrastructure for Computing (SNIC),
and Knut and Alice Wallenberg Foundation;
Australia {\textendash} Australian Research Council;
Canada {\textendash} Natural Sciences and Engineering Research Council of Canada,
Calcul Qu{\'e}bec, Compute Ontario, Canada Foundation for Innovation, WestGrid, and Compute Canada;
Denmark {\textendash} Villum Fonden and Carlsberg Foundation;
New Zealand {\textendash} Marsden Fund;
Japan {\textendash} Japan Society for Promotion of Science (JSPS)
and Institute for Global Prominent Research (IGPR) of Chiba University;
Korea {\textendash} National Research Foundation of Korea (NRF);
Switzerland {\textendash} Swiss National Science Foundation (SNSF);
United Kingdom {\textendash} Department of Physics, University of Oxford.

\end{document}